\input{aipcheck}

\documentclass[
    ,final            
  ]
  {aipproc}

\layoutstyle{6x9}

\begin{document}

\title{Interacting Convection Zones}


\author{L. J. Silvers}{
  address={DAMTP, University of Cambridge, Wilberforce Road,
  Cambridge, CB3 0WA, UK}
}
\author{M. R. E. Proctor}{ address={DAMTP, University of Cambridge,
  Wilberforce Road, Cambridge, CB3 0WA, UK}}

\begin{abstract}
We present results from simulations to examine how the separation between two convectively unstable layers affects their interaction. We show that two convectively ustable layers remain connected via the overshooting plumes even when they are separated by several pressure scale heights.

\end{abstract}

\maketitle


\section{Introduction}

Some main-sequence stars, such as A-type stars, have a complex internal
structure in which there are believed to exist multiple convection
zones \citep{TZLS,Kupka2}. Such a complicated internal structure arises because of the
change in the chemical composition as the distance from the centre of
the star is increased. The outermost convection zone, located just
below the photophere is belived to be caused by the partial ionization
of hydrogen and the single ionization of helium. The lower convection
zone is believed to result from the second ionization of helium and is
separated from the upper convection zone by a relatively narrow convectively stable
region \citep{LTZ,Kupka2}.

The presence of two convectively unstable zones in such
stars raises a number of interesting questions. However, in this short article
we choose to focus on the principal question concerning the degree of
intereaction between the convectivley unstable regions. Such an issue is
motivated  by the fact that in stars, without the presence of
rigid boundaries, ascending and decending convectively driven motions
overshoot the bottom of the layer that is convectively unstable. When
there are multiple convection zones, in close proximity, in stars this
overshooting behaviour leads to enhanced transport and communication
between the unstable layers. Further, due to the close proximity of
the unstable layers, fascinating dynamics may occur. For example,
overshooting plumes form each unstable layer could overshoot so far
that they continue through the stable region and pierce the other
convection zone, which would leed to transportation of 'contaminants'
directly from one convectively unstable region to the other \citep{TBCT1,TBCT2}. In this
article we will present some results from simulations to explore how increasing the separation affects the connection between the
layers.

It has been conjectured, from analytic theory, that two pressure scale
heights should be sufficient to `disconnect' the two convectively
unstable regions \citep{LTZ,TZLS}. Here, we will briefly show that this is an underestimate and
that to `disconnect' the two convectively unstable layers many more
scale heights of separation is required. A more detailed exploration
of this issue can be found in \citet{SP}.

\section[]{Model}
Here we examine the evolution of a compressible fluid in a Cartesian layer.
The governing equations are \citep{MPW}:
\begin{equation}
\frac{\partial{\rho}}{\partial{t}}+\mathbf{\nabla}.
\rho\textbf{u}=0, \nonumber
\end{equation}
\begin{equation}
\rho\left(\frac{\partial \textbf{u}}{\partial
t}+\textbf{u}.\mathbf{\nabla}\textbf{u} \right) = -\mathbf{\nabla} P
+ \theta (m+1)\rho \hat{\textbf{z}}+\sigma\kappa\mathbf{\nabla} . \rho
\mathbf{\tau},  \nonumber
\end{equation}
\begin{equation}
\frac{\partial{T}}{\partial{t}}+ \textbf{u}. \mathbf{\nabla} T  = -
(\gamma -1) T \mathbf{\nabla}. \textbf{u}  +
\frac{\kappa(\gamma-1)\sigma \tau^{2}}{2}
+\frac{\gamma \kappa}{\rho} \nabla^{2} T,
\end{equation}
 where $z$ is taken downward, $\theta$ is the dimensionless temperature difference across
the layer, $R_*$ is the gas constant, $m $ is the polytropic index,
$\kappa=K/d \rho_{0} c_{P} \sqrt(R_{*} T_{0})$ is the dimensionless
thermal diffusivity, $\gamma$ is the ratio of specific heats,
$\mathbf{\tau}$ is the stress tensor, $P = \rho T$ and $\sigma$ is the Prandtl number.These equations above are solved using a parallel hybrid
finite-difference/pseudo-spectral code; the most comprehensive
description of the code can be found in \citep{MPW}. 

To obtain a suitable basic state for this problem we allow the thermal profile to be non-linear and we take
\begin{eqnarray}
K  =  \frac{K_1}{2} \Big[1 &+& \frac{K_2+K_3}{K_1} + 
\frac{K_3}{K_1}\tanh\Big(\frac{z-\mathcal{D}+1}{\Delta}\Big)
\nonumber \\ &-& \Big(\frac{K_2}{K_1}\tanh\Big(\frac{z-\mathcal{D}+1}{\Delta}\Big)+1\Big)\tanh\big(\frac{z-1}{\Delta}\Big)\Big]
\end{eqnarray}
\noindent where $\Delta$ is the characteristic size of the
transition region between each of the layers. In this work  the
characteristic sizes of the transition regions are taken to be the same for
simplicity. The static density and temperature profiles are found by
solving the equations of hydrostatic balance. To this
static state, throughout the domain, random perturbations are
introduced, with amplitudes which lie within the interval [-0.05,0.05].

The aspect ratio for the computational domain in this study is
8:8:$\mathcal{D}$\footnote{Note that the more detailed exploration of
this topic in \citep{SP} is in a, smaller, 4:4:$\mathcal{D}$ domain. }, where $\mathcal{D}$ is the total depth of the
box, and the domain is assumed to be periodic in \textit{x} and
\textit{y}. The conditions on the upper and
lower boundaries are:

\begin{equation}
T= 1,\hspace{0.5cm} u_z=0, \hspace{0.5cm} \frac{\partial
u_x}{\partial z}= 0 \hspace{0.5cm} at \hspace{0.1cm} z=0. \nonumber 
\end{equation}
\begin{equation}
\frac{\partial T}{\partial z} = \theta ,\hspace{0.5cm} u_z=0,
\hspace{0.5cm} \frac{\partial u_x}{\partial z}= 0. \hspace{0.5cm} at
\hspace{0.1cm} z=\mathcal{D}. \nonumber
\end{equation}

The system we study has a large number of dimensionless parameters,
making it impractical to conduct a complete survey. Thus a number are
held fixed: $\sigma=1.0$, $m_1=m_3=1.0$, $\theta=10.0$, $\gamma=5/3$
a nd $R_a=1.7x10^5$. Note that in this paper we use a subscript 1 on quantities relating to the upper convection
zone. Similarly, subscript 2 will be used to denote quantities for
the convectively stable layer and 3 to denote quantities in the
lower convective zone.

The stiffness parameter, $S$, provides a useful measure of the
relative conductivities in this problem (for more detailed
discussion see \citet{TBCT1}). $S_2$ and $S_3$ are
related to the various polytropic indexes that appear in the problem
via the definition that $S_2=(m_2-m_{ad})/(m_{ad}-m_1)$ and $
S_3=(m_3-m_{ad})(m_{ad}-m_1)$.

\section{The Effect of Varying The Thickness of the Stable
Layer}\label{thick}

\begin{figure}
  \includegraphics[height=.2\textheight]{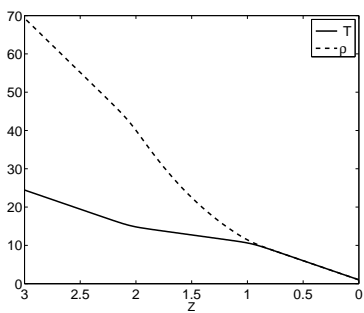}
  \caption{The initial temperature and density profiles when $\mathcal{D}=3.0$.}\label{fidback}
\end{figure}

 We begin by
considering the case where  all three zones have equal depth and so
$\mathcal{D}=3.0$.  The static state profiles
for this fiducial model are shown in Figure \ref{fidback}. Note that
for this fiducial case there are 1.60 pressure scale heights across the
convectively stable mid-layer, which is
rather less than the two pressure scale heights estimate suggested
by the earlier analytic theory as necessary for true separation. 


\begin{figure}
  \includegraphics[height=.2\textheight]{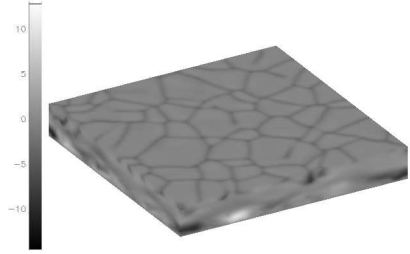}
  \caption{Plot at $t=15.02$ Sides of the box show the vertical momentum
  flux and the top of the box shows the vertical momentum flux near
  the top of the box for the case where $\mathcal{D}=3.0$ and $S_2=5.0$.}\label{fidevolved}
\end{figure}

The state shown in Figure \ref{fidback} is perturbed and allowed to
evolve to a statistically steady state as shown in Figure \ref{fidevolved}. As one might expect the convection is
noticeably different in the two convection zones. 
While Figure \ref{fidevolved} provides a useful picture of the
convective state it is difficult to gauge
the motions within the convectively stable region from such a
figure. To obtain a clearer picture of the potential interactions
between the two convectively unstable regions it is helpful to note
that, if the two layers are to be considered as `independent' from
each other, then there needs to be a region between the layers where the
velocity becomes very small. Therefore, to facilitate a clearer picture of
the degree of interaction of the layers we calculate the variation in $z$ of the horizontal averages of the modulus of
the $z$ component of momentum.

\begin{figure}
  \includegraphics[height=.2\textheight]{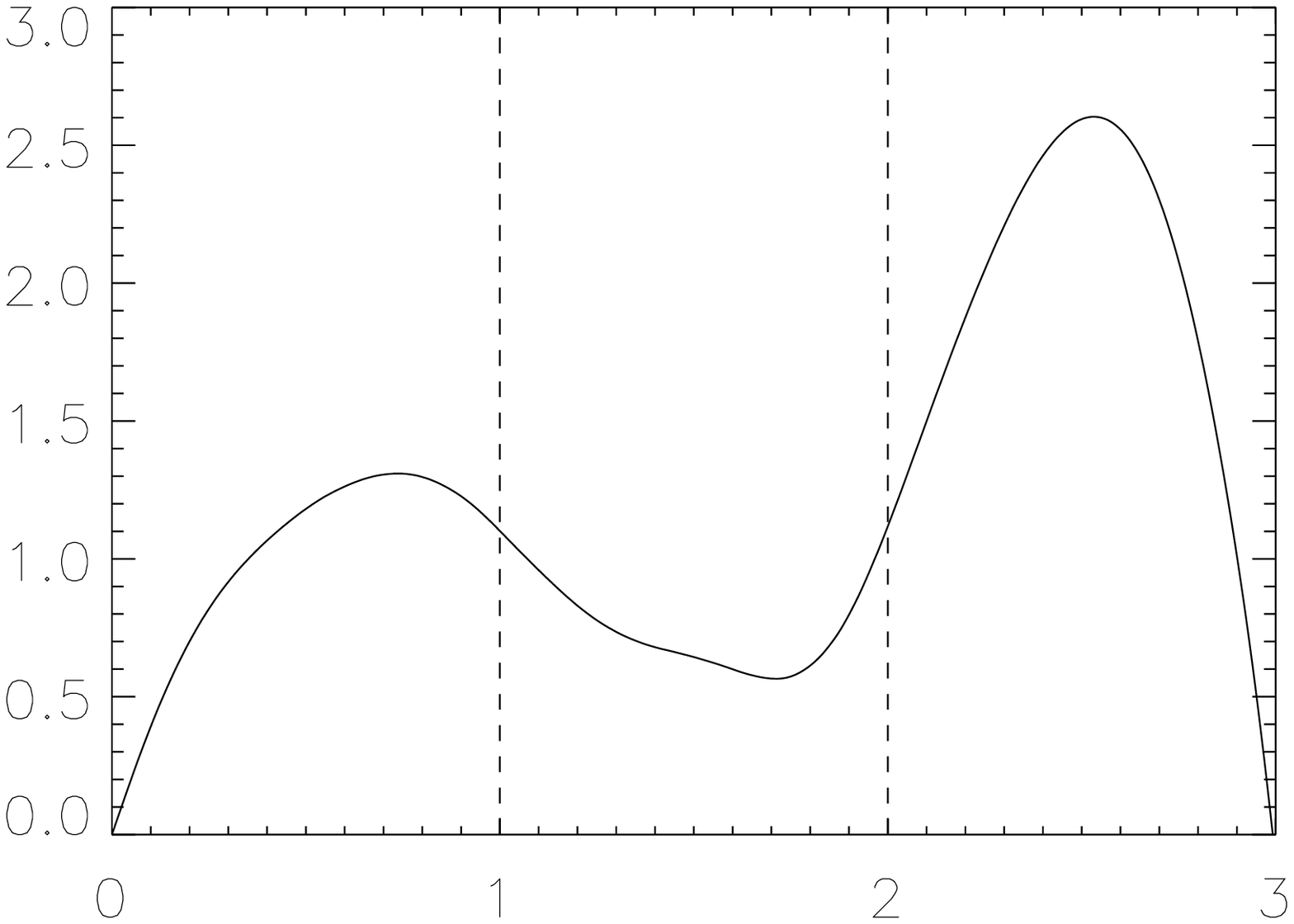}  \includegraphics[height=.2\textheight]{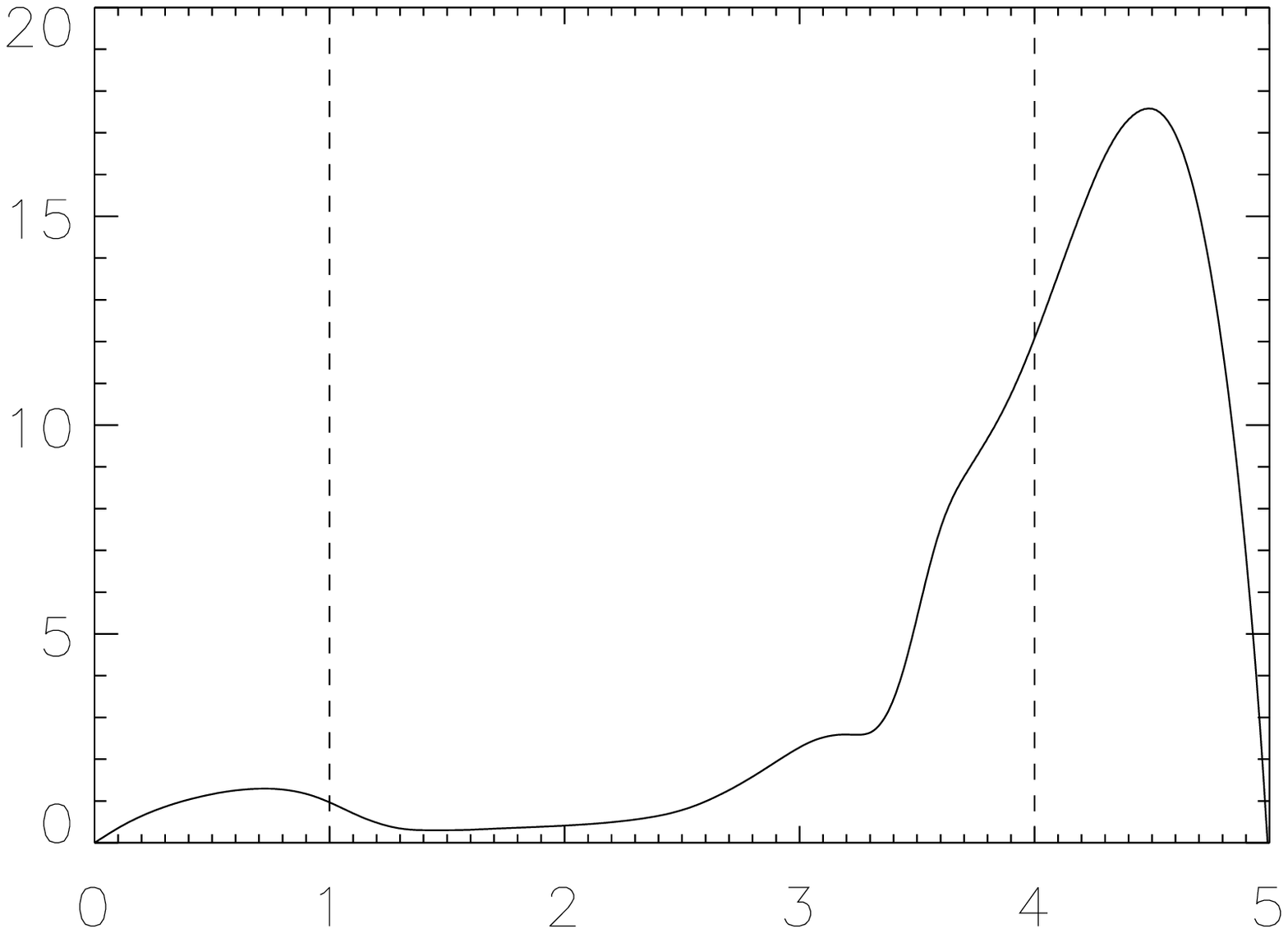}
  \caption{Left: Plot at $t=15.32$. The variation of the vertical
  component of velocity as a function of height at the centre of the
  box for the case where $\mathcal{D}=3.0$ and $S_2=5.0$. Right: Plot at $t=15.09$. The variation of the vertical component of velocity as a function of height at the centre of the box for the case where $\mathcal{D}=5.0$ and $S_2=5.0$.}\label{vv}
\end{figure}

Figure \ref{vv}(a) shows that while there is clearly more vertical motion in the
convectively unstable regions, as one might
anticipate,  there is still a non-negligible
vertical component of momentum in the middle of the box. Thus the two
convection zones are connected in this case and so there  is a conduit for mixing between the two convection
zones for this level of separation. Any reduction in the width of
the convectively stable region will decrease the number of pressure
scale heights of variation across the layer and increase the level
of interaction between the two unstable zones. Such vigorous motion in the convectively stable
layer was discussed in the context of the downward directed hexagon case of Latour \textit{et al.} and is further confirmed
by a plot of the vertical component of velocity as shown in
\ref{vv}(a).

The most obvious way to limit the interaction of the unstable
regions is to increase the layer depth, $\mathcal{D}$, and so increase the
width of the stable region if the convectively unstable regions have
the same height.  We have gradually increased the layer depth up to
5.0. This corresponds to a pressure scale height variation of 3.77
across the stable layer, which is abover the estimate suggested by the
earlier analytic theory as the transition between connected and
unconnected convection layers.

As for the case when $\mathcal{D}=3.0$, we consider the variation of
the planar average vertical component of momentum throughout the
domain in the established statistically steady state, for each of
the increased box heights. Figures \ref{vv}(b) show this
quantity for each case when $\mathcal{D}=5.0$. This figures clearly illustrates that, while there is a
significant reduction in vertical motion in the convectively stable
region, there is a non negligible vertical component of momentum
throughout the box even for a separation of 3.77 pressure scale heights.

\section{Conclusions}

In this article we have shown that there is no point
at which the vertical component of momentum becomes very small even when the
number of pressure scale heights of separation between the two
convectively unstable layers is as large as 3.77. Thus
there is a route for mixing of passive and dynamic quantities
between the two convectively unstable regions in all the cases we
considered.  Only for a substantial
increase in the box height will the
vertical component of momentum fall to a very small value at one plane in the box. To find the minimum
box height for this to occur would be extremely
computationally expensive and also uninteresting 
physically. In A-type main sequence stars the separation
between the unstable layers does not extend past a couple of scale
heights. Therefore, on the basis of the present work we can conclude
that in A-type stars there is a clear connection between the
convectively unstable zones that lie immediately below the stellar
photosphere.

One of the major objectives of this work  is to provide a solid
hydrodynamical basis on which more complex models can be constructed to 
understand fully the dynamics that occur below the surface of A-type
stars. With such a simple model we are not yet in a position to address important secondary questions such as the influence of the observed chemical anomalies (see discussions in
\citet{{Michaud1970}} or \citet{{Vauclair1982}}  for
more details). However, we believe our model provides a platform on which we can build, so as to address
the effects of rotation, magnetic fields and other
issues related to the dynamics in these stars and will be the subject of future work.


\begin{theacknowledgments}
 LJS wishes to thank the Department of Applied Mathematics and
Theoretical Physics at the University of Cambridge for the award of
a Crighton fellowship for partial support of this research. We would
like to thank Paul Bushby, Steve Houghton and Douglas Gough
for helpful discussions.

\end{theacknowledgments}

\end{document}